\documentclass[12pt]{iopart}

\usepackage{iopams}
\usepackage{graphicx}

\begin{document}

\title[Optical phase dynamics...]{Optical phase dynamics in mutually coupled diode laser systems exhibiting power synchronization}

\author{Vishwa Pal$^1$, Awadhesh Prasad$^2$ and R Ghosh$^1$}

\address{$^1$ School of Physical Sciences, Jawaharlal Nehru University, New Delhi 110067, India}
\address{$^2$ Department of Physics and Astrophysics, University of Delhi, Delhi 110007, India}
\ead{rghosh.jnu@gmail.com}

\begin{abstract}
We probe the physical mechanism behind the known phenomenon of power synchronization of two diode lasers that are mutually coupled via their delayed optical fields. In a diode laser, the amplitude and the phase of the optical field are coupled by the so-called linewidth enhancement factor, $\alpha$. In this work, we explore the role of optical phases of the electric fields in amplitude (and hence power) synchronization through $\alpha$ in such mutually delay-coupled diode laser systems. Our numerical results show that the synchronization of optical phases drives the powers of lasers to synchronized death regimes. We also find that as $\alpha$ varies for different diode lasers, the system goes through a sequence of in-phase amplitude-death states. Within the windows between successive amplitude-death regions, the cross-correlation between the field amplitudes exhibits a universal power-law behaviour with respect to $\alpha$.
\end{abstract}

(Some figures in this article are in colour only in the electronic version)



\submitto{\JPB}
\maketitle

\section{Introduction}
Synchronization in coupled oscillators has been the subject of much attention because of its fundamental importance in many areas of science and technology, such as in the dynamics of lasers \cite{roy93}, electronic circuits \cite{pecora94}, biological systems \cite{kuramoto95}, and its application especially in communication \cite{carroll90, rahman90, thornburg94, shimizu94, roy98, porte98}. Synchronization is known to be caused by the interaction between two nonlinear oscillators as a result of coupling, and, depending on the coupling strength and the time-delay in coupling, various features of synchronization emerge. The system of two coupled lasers is known to be an excellent, experimentally realizable example of coupled nonlinear oscillators, and synchronization in this context refers to the phenomenon in which the intensities of two lasers have a well-defined relation at all times. Different types of relations between laser intensities describe different types of synchronization \cite{sync-book}, e.g., {\it generalized} synchronization when intensities are functionally related to each other, {\it complete} synchronization when intensities of the lasers become identical (as a particular case of the generalized one), etc. There is another type called {\it phase} synchronization, when intensities of the lasers are uncorrelated but the phase difference of their oscillations remains bounded. This phase synchronization is usually defined as locking of the phases $\theta_{1,2}$, $|p\theta_{1} - q\theta_{2}| <$ constant \cite{kurths96}, where $p$ and $q$ are integers. The phase here indicates the phase of oscillations of the laser intensity, and not the optical phase of the electrical field. The phase synchronization of the oscillations of the laser intensities is studied by using the analytical signal concept proposed by Gabor \cite{gabor46, kurths96, roy01}. There exists a large body of work on synchronization and communication with {\it chaotic} laser systems -- see Ref.\,\cite{roy05} for a review.

For the study of the synchronization phenomenon, mutually delay-coupled diode lasers are suitable candidates because of the compactness, low cost, and durability of diode lasers. Different aspects of the complex dynamics of mutually delay-coupled diode laser system have been probed -- see Refs.\,\cite{rogister04, erzgraber05} and references therein. The system provides a simple and powerful tool to unveil the collective behaviour within a wide range of control parameter space, spanned by the coupling strength and the time-delay in coupling. Among the collective behaviour, amplitude death can occur in which two identical or non-identical coupled oscillators drive each other to a fixed point and stop the oscillations \cite{eli85}. We have recently done experimental and theoretical studies of amplitude death and phase synchronization of powers in a mutually delay-coupled diode laser system \cite{kumar08, kumar09}; however, to the best of our knowledge, the underlying physical mechanism of the power synchronization is not yet understood clearly. Given the importance of the system for fundamental studies as well as applications, we wish to extend our work to the understanding of the role, if any, of the optical phases of the laser fields in the process of power synchronization, for fixed values of coupling strengths and delays in the region of interest.

This work can be put in the context of general research involving control of nonlinear dynamical systems using time-delayed feedback -- see Ref.\,\cite{hinz11} and references therein. In the systems studied in optics, {\it passive} feedback from mirrors or external cavities or resonators has been used for this purpose. A notable mention is the work by Schikora {\it et al.} \cite{schikora06} for control of unstable steady-states in a semiconductor laser, which pointed out the role of the optical phase in the feedback control scheme. Our interest is in the understanding of the dynamical features of delayed coupling between two {\it active} devices of diode lasers, in the {\it periodic} regime of system parameters. In a diode laser, the active material has a highly asymmetric gain profile. This bears consequences to the refractive index (real part of the susceptibility), which can be related to the gain (imaginary part of the susceptibility) by the Kramers-Kronig relations \cite{henry81}. The increase of the gain in diode lasers by increasing the carrier density leads to a decrease of the refractive index. The strength of the coupling between gain and refractive index is described by a parameter $\alpha$, known as amplitude-phase coupling or linewidth enhancement factor \cite{henry86,henry82}. The $\alpha$-factor influences several fundamental aspects of all semiconductor lasers, such as the linewidth, the chirp under current modulation, the mode stability, the occurrence of filamentation in broad-area devices. The dynamics of semiconductor lasers is greatly influenced by the $\alpha$-factor, and its role has been specifically probed for dynamical effects such as instability enhancement in semiconductor lasers with delayed feedback \cite{wada94} and injection-locking in semiconductor laser amplifiers \cite{taraprasad96}.

In a coupled diode laser system, an amplitude fluctuation in one laser leads to a carrier density fluctuation, and through $\alpha$, a phase fluctuation in the same laser. A change in the relative phase leads to an amplitude change in the second laser and an accompanying change in its carrier density. Thus a natural question arises: How does the phase relation between laser fields emerge in the complex dynamics? Does it have a bearing on the amplitude synchronization of the lasers? What is the effect of tuning the $\alpha$-factor on the correlated amplitude dynamics? In this work, using the standard theoretical model for the system, we explore numerically the role of optical phases of the electric fields in power synchronization in a mutually delay-coupled diode laser system.

The measured output powers of the lasers, $P \equiv |E|^2 = A^2$ of course do not explicitly depend on the phases $\phi$ of the optical fields $E$, but there is indeed a connection between the field amplitudes $A$ and the field phases $\phi$ through $\alpha$. Our numerical results show that the synchronization of optical phases drives the powers of lasers to the synchronized death regime for a fixed $\alpha$. Moreover, the cross-correlation measure between the field amplitudes shows that the system exhibits a sequence of in-phase amplitude-death regimes as $\alpha$ is varied for different diode lasers. In between successive in-phase amplitude-death regions, the cross-correlation coefficient shows a scaling behaviour with respect to $\alpha$. It exhibits a power law, representing the transition from the in-phase amplitude-death to anti-phase periodic oscillations, and vice versa.

In this paper, the theoretical model and numerical details to describe the two mutually delay-coupled diode lasers are presented in section $2$. Section $3$ presents the temporal dynamics of amplitudes and phases of the optical fields in such coupled lasers exhibiting synchronized death. The correlation measure is used in section $4$ to explore the behaviour of synchronization over a large range of the amplitude-phase coupling $\alpha$. This section also contains the scaling of the cross-correlation with respect to $\alpha$ within the windows between successive in-phase amplitude death regions. The summary of the work is presented in section $5$.

\section{Theoretical model}
A single diode laser subjected to optical feedback from an external cavity can be modelled by a set of fundamental delay-differential equations which are known as Lang-Kobayashi (LK) equations \cite{lang80}. These equations describe the time evolutions of the complex electric field $E(t)$ of a single longitudinal mode and the carrier density $N(t)$ (with the threshold value subtracted out) averaged over the laser medium.
\begin{figure}[htbp]
\begin{center}
\scalebox{0.55}{\includegraphics{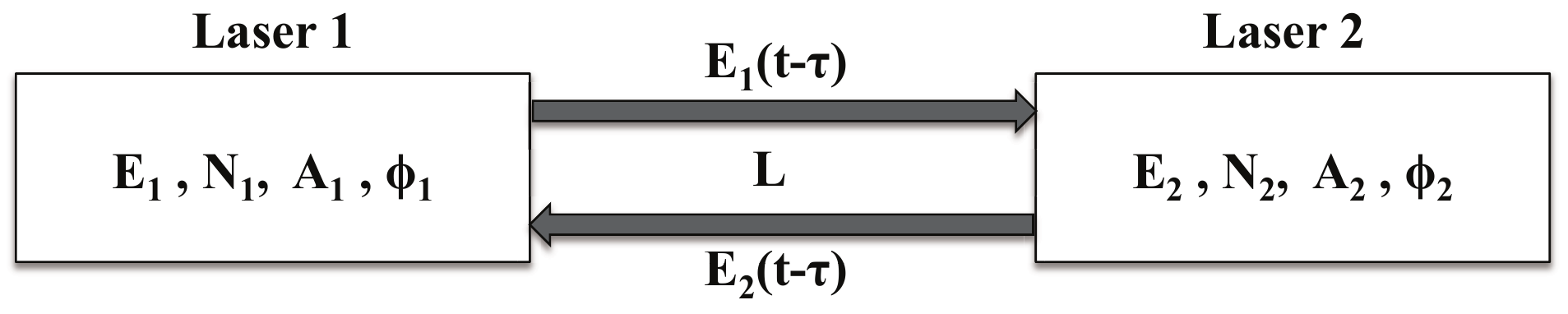}}
\end{center}
\caption{\scriptsize Two diode lasers coupled in face-to-face configuration (schematic).}
\label{coupledlasers}
\end{figure}
In order to analyse the behaviour of two diode lasers coupled in face-to-face configuration (as shown in figure \ref{coupledlasers}), the LK equations can be written in a standard normalized form \cite{kumar08, kumar09}:
\begin{eqnarray}
\frac{dE_{1}}{dt}&=&(1+i\alpha)N_{1}(t)E_{1}(t)+\eta ~\mathrm{e}^{-i\omega_{2}\tau}E_{2}(t-\tau),\label{eq1}\\
T\frac{dN_{1}}{dt}&=&J_{1}-N_{1}-\left[2N_{1}+1\right]\left| E_{1}(t)\right|^{2},\label{eq2}\\
\frac{dE_{2}}{dt}&=&(1+i\alpha)N_{2}(t)E_{2}(t)+\eta ~\mathrm{e}^{-i\omega_{1}\tau}E_{1}(t-\tau),\label{eq3}\\
T\frac{dN_{2}}{dt}&=&J_{2}-N_{2}-\left[2N_{2}+1\right]\left| E_{2}(t)\right|^{2},\label{eq4}
\end{eqnarray}
where $\eta$ is the coupling strength, i.e., the fraction of light of one laser injected into the other laser and vice versa, $J$'s are the injected constant current densities (with the threshold value subtracted out), $T$ is the ratio of the carrier lifetime to the photon lifetime, $\alpha$ is the linewidth enhancement factor as before, and $\tau$ is the time taken by the light to cover the distance between the lasers. $\omega_{1,2}$ are the optical angular frequencies of the solitary lasers 1 and 2. $E_{1,2}(t-\tau)$ are the fields delayed by one coupling time $\tau=L/c$ and $\omega_{1,2}\tau$ are the phase mismatches. For simplicity, we have taken two identical diode lasers for which the detuning between them is assumed to be zero. In order to keep the model simple, we have ignored noise sources. The complex fields of the lasers can be written in terms of amplitude and phase parts as
\begin{equation}
E_{i}=A_{i}~\mathrm{e}^{-i\phi_{i}},\label{eq5}
\end{equation}
where $A_{i}$ and $\phi_{i}$ are the amplitude and the phase of the optical field of laser $i$ ($i$ = 1, 2), respectively. Using (\ref{eq5}) in (\ref{eq1})-(\ref{eq4}), we get
\begin{eqnarray}
\frac{dA_{1}}{dt}&=&N_{1}(t)A_{1}(t)+\eta A_{2}(t-\tau)\cos\left[\phi_{1}(t)-\phi_{2}(t-\tau)-\omega_{2}\tau \right],\label{eq6}\\
T\frac{dN_{1}}{dt}&=&J_{1}-N_{1}-\left[2N_{1}+1\right]\left| A_{1}(t)\right|^{2},\label{eq7}\\
\frac{d\phi_{1}}{dt}&=&-\alpha N_{1}(t)-\eta \frac{A_{2}(t-\tau)}{A_{1}(t)}\sin\left[\phi_{1}(t)-\phi_{2}(t-\tau)-\omega_{2}\tau \right],\label{eq8}\\
\frac{dA_{2}}{dt}&=&N_{2}(t)A_{2}(t)+\eta A_{1}(t-\tau)\cos\left[\phi_{2}(t)-\phi_{1}(t-\tau)-\omega_{1}\tau \right],\label{eq9}\\
T\frac{dN_{2}}{dt}&=&J_{2}-N_{2}-\left[2N_{2}+1\right]\left| A_{2}(t)\right|^{2},\label{eq10}\\
\frac{d\phi_{2}}{dt}&=&-\alpha N_{2}(t)-\eta \frac{A_{1}(t-\tau)}{A_{2}(t)}\sin\left[\phi_{2}(t)-\phi_{1}(t-\tau)-\omega_{1}\tau \right].\label{eq11}
\end{eqnarray}
Note that the measured output powers of the lasers, $P_i \equiv |E_i|^2 = A_i^2$ do not explicitly depend on the phases $\phi_i$ of the optical fields. We still wish to probe the role of phases of optical fields in the power synchronization of two mutually delay-coupled diode lasers as $\alpha$ connects the phases and the amplitudes of the optical fields. Numerical integration of the above equations is done using Runge-Kutta fourth-order scheme with a step size $= \tau /n$, where $n=1000$ is chosen based on the accuracy criteria. The dimensionless parameters are taken as $J_{1,2}=0.165$, and $T=1000$ \cite{kumar08, kumar09}. It is found that the phase mismatch does not influence the results qualitatively and thus we keep $\omega_{1,2}\tau=-1$ (mod $2\pi$) \cite{mandel04}. Different kinds of lasers have different ranges of the linewidth enhancement factor $\alpha$, e.g., for gas lasers, $\alpha$ is $0$; for quantum dot lasers, $\alpha$ is $1.5$ to $3$ \cite{muszalski04, ukhanov04}, and for conventional diode lasers, $\alpha$ is typically $2$ to $6$ \cite{sang89}.

\section{Synchronized death}
We analyze the temporal behaviours of field amplitudes and optical phase difference in the amplitude-death regime, where the two mutually delay-coupled oscillators drive each other to a fixed point and stop the oscillations, as seen in \cite{kumar08, kumar09}. The temporal behaviour of amplitudes and phases of the optical fields are obtained numerically by integrating equations (\ref{eq6})-(\ref{eq11}). In order to remove the initial transients, it is sufficient to discard the first 50000 data points before performing the synchronization study. We first choose a typical value of $\alpha=5.6$ corresponding to a diode laser, as in \cite{kumar08}.
\begin{figure}[htbp]
\begin{center}
\scalebox{0.55}{\includegraphics{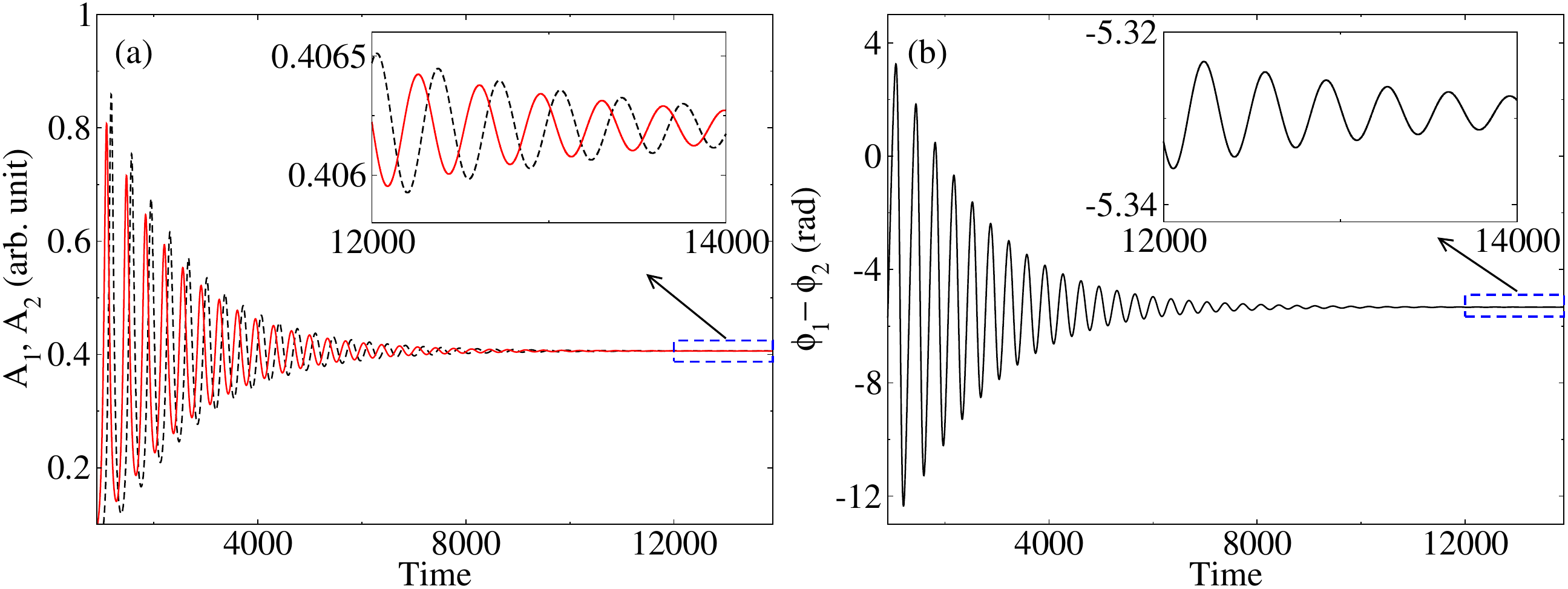}}
\end{center}
\caption{\scriptsize The time (units of cavity photon life-time) series of (a) the laser field amplitudes, $A_{1}$ and $A_{2}$ (continuous and dashed lines), and (b) the optical phase difference, $\phi_{1}-\phi_{2}$, for {\it uncoupled} lasers ($\eta$ = 0) at $\alpha=5.6$, with initial conditions $A_1(0)\ne A_2(0)$, $N_1(0) \ne N_2(0)$, and $\phi_1(0) \ne \phi_2(0)$. The initial transients are not shown. Insets show magnified temporal behaviours in the marked regions.}
\label{temp_dyna1}
\end{figure}

\begin{figure}[htbp]
\begin{center}
\scalebox{0.55}{\includegraphics{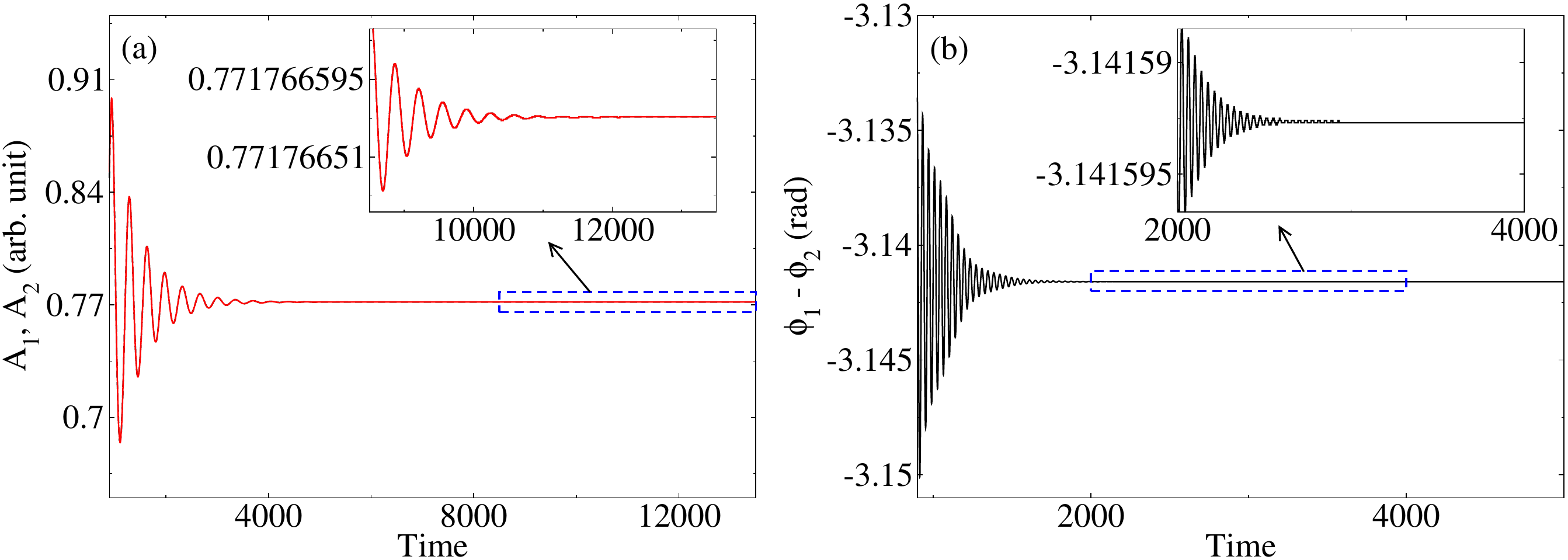}}
\end{center}
\caption{\scriptsize The same as figure \ref{temp_dyna1}, but for the {\it coupled} lasers with a coupling strength $\eta$=0.2, time delay $\tau=14$ and $\alpha=5.6$.}
\label{temp_dyna2}
\end{figure}

From figure \ref{temp_dyna1}, it is clear that when the coupling $\eta$ between the lasers is zero, the laser field amplitudes (figure \ref{temp_dyna1}(a)) and their optical phases (phase difference between the laser fields) (figure \ref{temp_dyna1}(b)) do not synchronize to some constant values in a practical time scale. These are clearly shown in the corresponding insets, where the oscillations around 0.4062 (arbitrary unit) and $-5.3292$ rad, respectively, are still present even after a long time span. In this case, an amplitude change in one laser causes (through $\alpha$) a phase change in the same laser, but the change in the relative phase does not lead to an amplitude change in the second laser. Thus due to the lack of this interaction, the amplitudes and the phases of the two laser fields do not synchronize to constant values. As the strength of coupling $\eta$ between lasers is switched on to a moderate value, e.g., $\eta$ = 0.2 and $\tau$ = 14 (in units of cavity photon lifetime), the field amplitudes and optical phase difference synchronize to constant values of $0.7717$ (arbitrary unit) and $-3.1415$ rad, as shown in figures \ref{temp_dyna2}(a) and \ref{temp_dyna2}(b), respectively. In the case when the coupling strength $\eta$ between the lasers is non-zero, the optical phase difference of the fields approaches a constant value much earlier than the field amplitudes. This implies that the phase synchronization precedes amplitude-death synchronization, and it drives the field amplitudes to the synchronized death state.

\begin{figure}[htbp]
\begin{center}
\scalebox{0.49}{\includegraphics{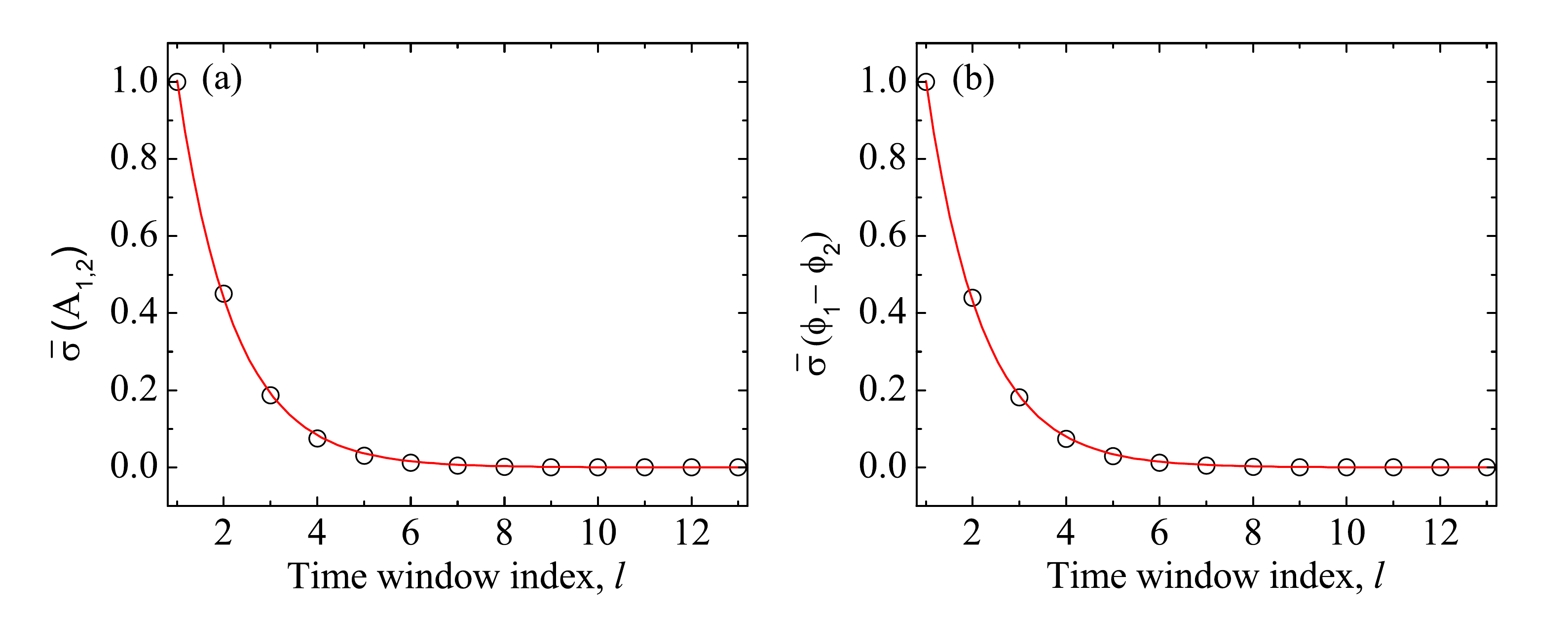}}
\end{center}
\caption{\scriptsize Scaled variance in (a) field amplitudes, $A_{1,2}$, and (b) optical phase difference, $\phi_1 - \phi_2$, versus time-window index, $l$, for the {\it uncoupled} system ($\eta$ = 0) at $\alpha$ = 5.6, using a time-window duration, $\Delta t$ = 700 units. Circles are the numerical data, and continuous lines are fitted curves using equation (\ref{eq13}) with (a) $a=2.296$, $\gamma~\Delta t = 0.828$ units, and (b) $a=2.330$, $\gamma~\Delta t = 0.844$ units.}
\label{osc_decay1}
\end{figure}

\begin{figure}[htbp]
\begin{center}
\scalebox{0.49}{\includegraphics{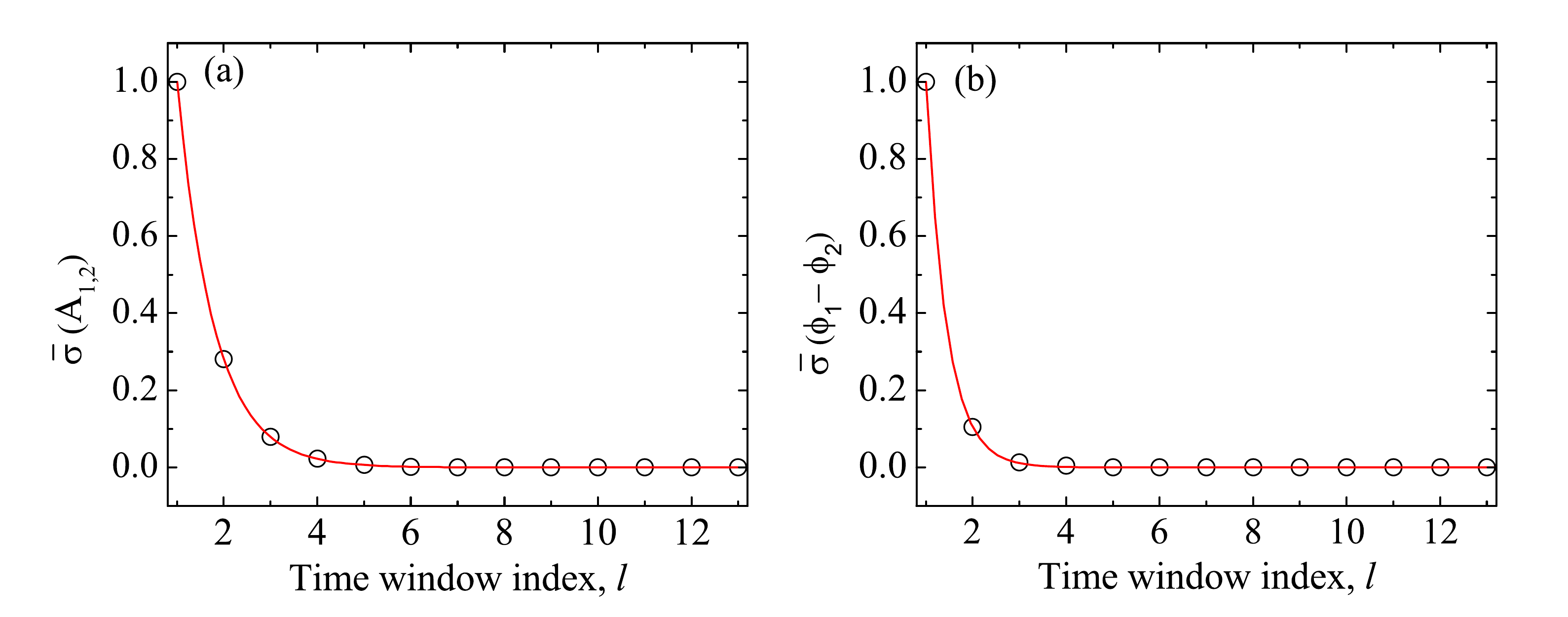}}
\end{center}
\caption{\scriptsize Scaled variance versus time-window index, $l$, for (a) field amplitudes, $A_{1,2}$, using a time-window duration $\Delta t$ = 350 units, and (b) optical phase difference, $\phi_1 - \phi_2$, using a time-window duration, $\Delta t$ = 112 units, for a {\it coupled} system with a time delay $\tau=14$, a coupling strength $\eta=0.2$ and an $\alpha=5.6$. Circles are the numerical data, and continuous lines are fitted curves using equation (\ref{eq13}) with (a) $a=3.562$, $\gamma~\Delta t = 1.270$, and (b) $a=9.541$, $\gamma~\Delta t = 2.256$.}
\label{osc_decay2}
\end{figure}

In order to quantify the decay of oscillations in laser field amplitudes and phase difference stated above, we analyze the variance \cite{atay10} as the measure of oscillations in the laser field amplitudes, $A_1,A_2$ or the phase difference, $\phi_1-\phi_2$, defined as
\begin{equation}
 \sigma(Q)=\Delta Q^{2}=\left\langle \left( Q-\left\langle Q\right\rangle \right)^{2}\right\rangle, \label{eq12}
\end{equation}
where $Q = A_{1,2}~ \mathrm{or} ~\phi_1 - \phi_2$, and $\langle \ldots \rangle$ denotes time-averaging. This can be calculated by splitting the time series of field amplitudes and phase difference into non-overlapping time windows of duration $\Delta t$. After removing the transients, we calculate $\sigma (Q)$ using a suitable $\Delta t$. We proceed to calculate the same for successive time windows ($l = 1, 2, \ldots$) of the same duration $\Delta t$, and stop when $\sigma < (0.001 ~ \times$ the value in the first time-window). The values of $\sigma (Q)$ are normalized to $\bar{\sigma}(Q)$, taking the maximum  as $\bar{\sigma}(Q)$ = 1 at the first ($l$ = 1) time-window, and these data points are marked as circles in figures \ref{osc_decay1} and \ref{osc_decay2}, for the uncoupled ($\eta$ = 0) and coupled ($\eta=0.2$, $\tau=14$) cases, respectively. For the uncoupled case with $\alpha$ = 5.6, we choose $\Delta t$ = 700 (units of cavity photon life-time). For the coupled lasers, the oscillations in field amplitudes and optical phase difference decay rather fast. Thus in order to generate sufficient number of data points, we split the time series of field amplitudes and optical phase difference into non-overlapping time windows of duration $\Delta t$ of 350 and 112 (units of cavity photon life-time), respectively.

The figures suggest that these oscillations decay exponentially, and can be fitted with the following expression:
\begin{equation}
\bar{\sigma}(Q)=a~\mathrm{e}^{-\gamma~l~\Delta t}, \label{eq13}
\end{equation}
where $a$ is a constant, $\gamma$ is the rate of decay per $\Delta t$, and $l$ is the time-window index. From the fits in the uncoupled case ($\eta=0$), shown in figures \ref{osc_decay1}(a) and \ref{osc_decay1}(b), the decay rates ($\gamma~\Delta t$) of oscillations in $A_{1,2}$ and $\phi_{1}-\phi_{2}$ are found to be $0.828$ and $0.844$ per $l$, respectively. Thus, the oscillations in laser field amplitudes and in phase difference decay more or less with the same rate in the uncoupled case.

Now, for the coupled system with coupling strength $\eta=0.2$ and $\tau$ = 14 units with $\alpha=5.6$, by fitting equation (\ref{eq13}), we find the decay rates ($\gamma~\Delta t$) of oscillations in $A_{1,2}$ and $\phi_{1}-\phi_{2}$ to be $1.270$ and $2.256$ per $l$, respectively. Converting these decay rates per the same units of time-window duration as taken for the uncoupled case, we obtain $(1.270 \times 700/350) = 2.54$ and $(2.256 \times 700/112) = 14.1$ for the two cases, respectively, demonstrating that the oscillations in both field amplitudes and phase-difference decay much faster than those in the uncoupled case, and for the coupled system, the field-amplitude oscillations decay much slower than the oscillations in the optical phase difference between the laser fields. Thus, for the coupled system, the phase difference synchronizes fast, and it then leads to the synchronized death in field amplitudes. This is a central result of the paper, providing an insight into the mechanism of power synchronization in a mutually coupled diode laser system.

\section{Correlation coefficient}
\begin{figure}[htbp]
\begin{center}
\scalebox{0.55}{\includegraphics{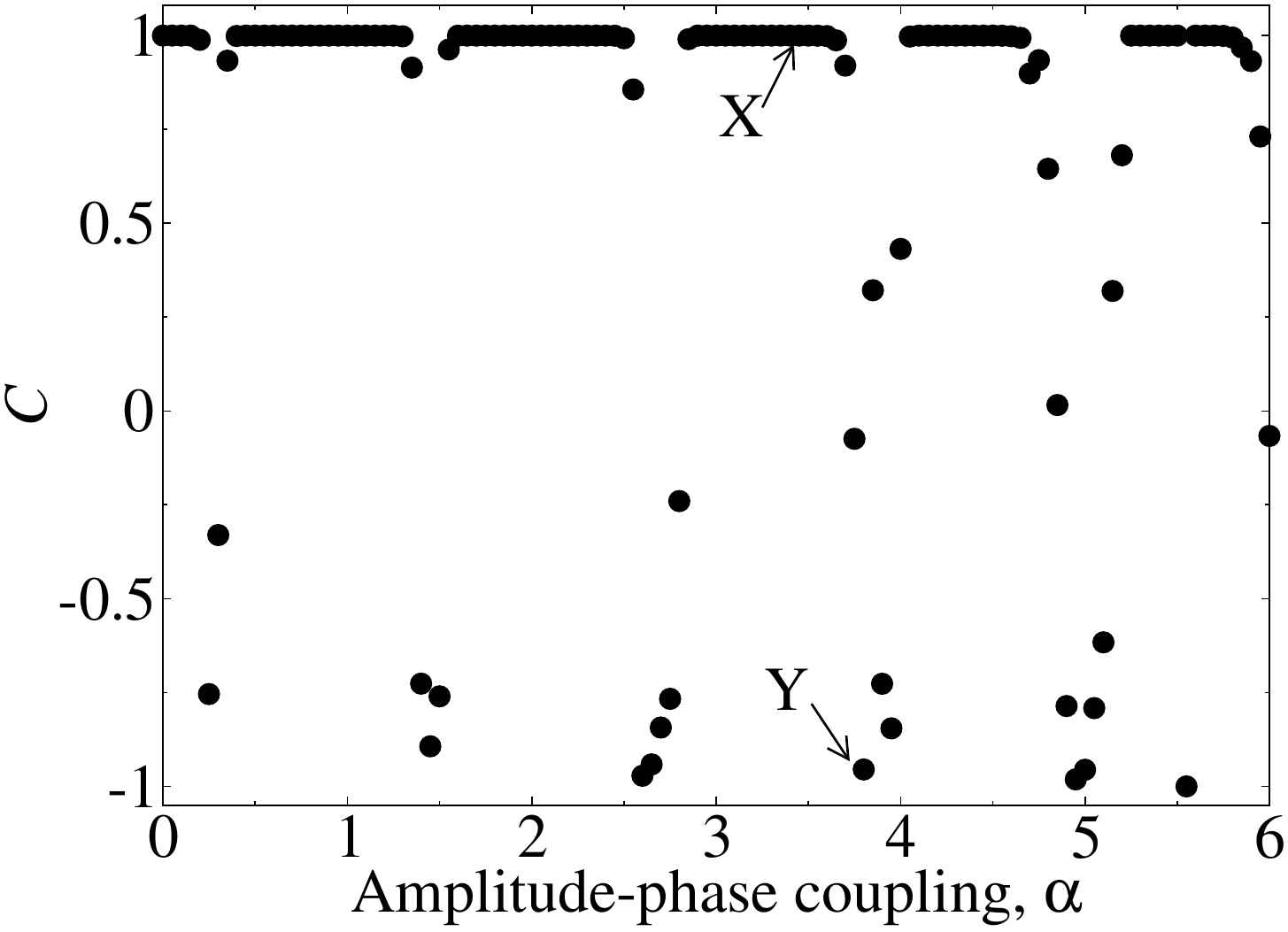}}
\end{center}
\caption{\scriptsize Plot of cross-correlation $C$, given by equation (\ref{eq14}), versus amplitude-phase coupling $\alpha$ for a time delay $\tau=14$ (in units of cavity photon life time) and a coupling strength $\eta=0.2$. The different regimes are marked as X: in-phase amplitude death, and Y: anti-phase periodic oscillations.}
\label{corr1}
\end{figure}

The amplitude-phase coupling $\alpha$ describes the coupling between the real and imaginary parts of the susceptibility, and is given by the ratio of their derivatives with respect to the carrier density. Its value is known to depend on the carrier concentration, photon energy and operating temperature \cite{osinski87}. For gain-guided and low-dimensional lasers (quantum wells and quantum wires), the value of $\alpha$ can in fact be controlled by the design of the device structure. In order to analyse the role of $\alpha$ in the synchronization of mutually delay-coupled diode lasers of different kinds, we use a normalized cross-correlation function defined as
\begin{equation}
C=\frac{\langle(A_{1}(t)-\langle A_{1}(t)\rangle)(A_{2}(t)-\langle A_{2}(t)\rangle)\rangle}{\sqrt{\langle (A_{1}(t)-\langle A_{1}(t)\rangle)^{2}\rangle \langle (A_{2}(t)-\langle A_{2}(t)\rangle)^{2}\rangle}}.\label{eq14}
\end{equation}
We probe the characteristic features of $C$ as the amplitude-phase coupling parameter $\alpha$ is tuned, and also mark the amplitude and phase dynamics at each distinct ($\alpha$, $C$) point. The variation of $C$ with $\alpha$, in a typical range for diode lasers, is shown in figure \ref{corr1}, using the numerical solutions of equations (\ref{eq6})-(\ref{eq11}). The various flat regions shown in figure \ref{corr1} have a cross-correlation coefficient $\approx 0.99$. We have checked the dynamics and the cross-correlation values for higher values of the coupling $\eta$ and the delay time $\tau$. It has been seen that with higher values of $\eta$, the flat regions of amplitude-death shrink to smaller ranges of $\alpha$, and with higher values of $\tau$, more transitions take place in the same range of $\alpha$, the qualitative features remaining the same.

\begin{figure}[htbp]
\begin{center}
\scalebox{0.55}{\includegraphics{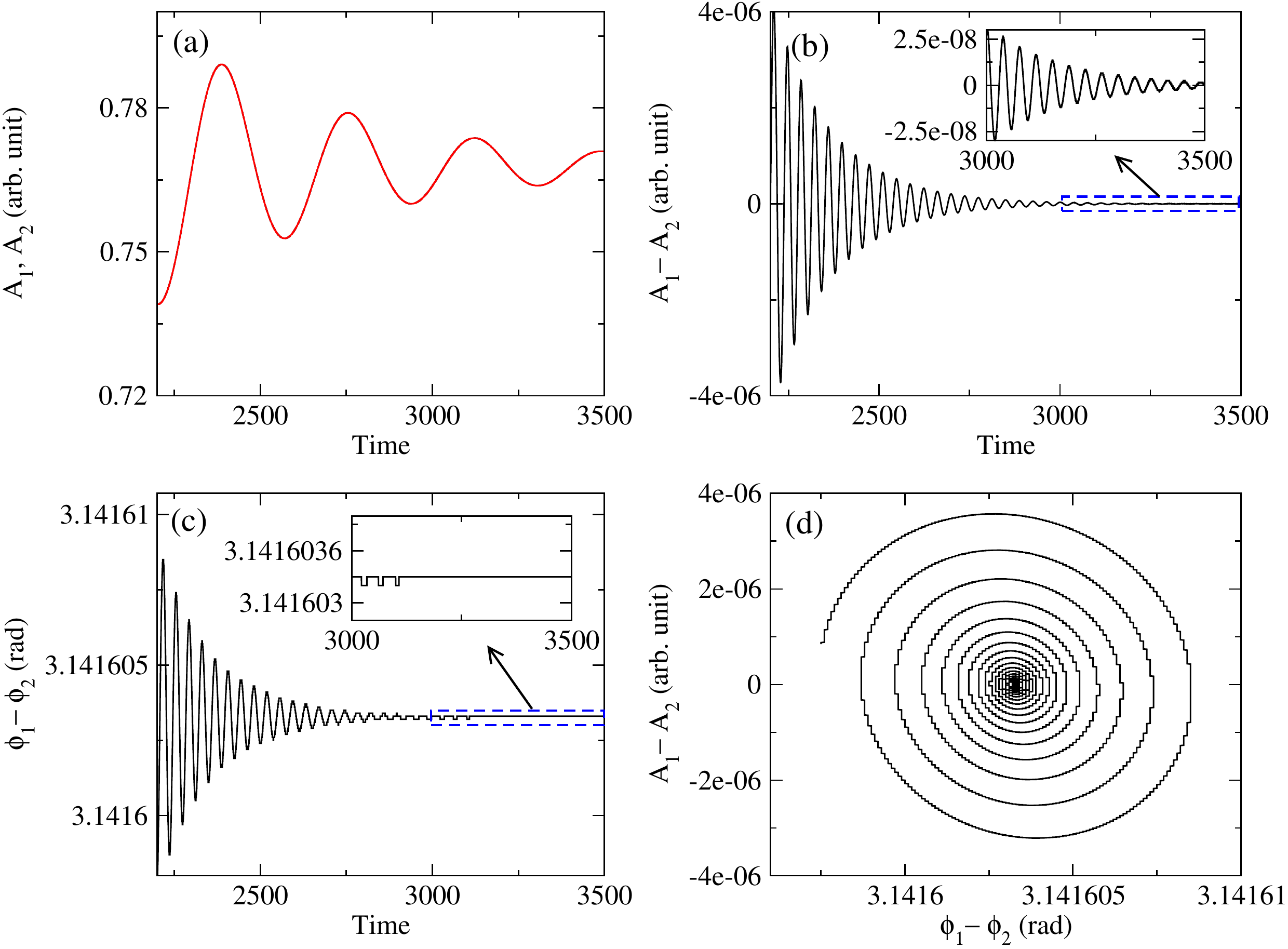}}
\end{center}
\caption{\scriptsize Plots of (a) laser field amplitudes $A_{1}$ and $A_{2}$ versus time (in units of cavity photon lifetime), (b) magnified $A_1 - A_2$ versus the same time (with the inset showing further magnified behaviour in the marked region), (c) $\phi_1 - \phi_2$ versus the same time (with the inset showing $\phi_1 - \phi_2$ = constant even in a magnified scale in the marked region), and (d) equal-time $A_1 - A_2$ versus $\phi_1 - \phi_2$ in the same time span, showing synchronization in the death state at $\alpha=3.40$, for a time delay $\tau=14$ units and a coupling strength $\eta=0.2$.}
\label{sync-death}
\end{figure}

In figure \ref{corr1}, the symbol X points to an in-phase amplitude-death region, while the symbol Y marks a region of anti-phase periodic oscillations. Corresponding to an in-phase amplitude-death state at $\alpha$ = 3.4 in figure \ref{corr1}, we show the synchronization dynamics in figure \ref{sync-death}. For all such in-phase amplitude-death regions, field-phase synchronization precedes amplitude-death synchronization (as discussed in the previous section). This is again seen in figure \ref{sync-death}, where the field-phase synchronization is established quite early, with the phase-difference $\phi_1 - \phi_2$ getting firmly locked at a constant value, as shown in figure \ref{sync-death}(c). $A_1$ and $A_2$ values are not exactly synchronized at these times, but the dynamics is shown to lead to the synchronized amplitude-death state at later times in figure \ref{sync-death}(b). The combined synchronization in the form of equal-time plot of the field amplitude-difference versus the field phase-difference, in the same time span, is presented in figure \ref{sync-death}(d).

\begin{figure}[htbp]
\begin{center}
\scalebox{0.55}{\includegraphics{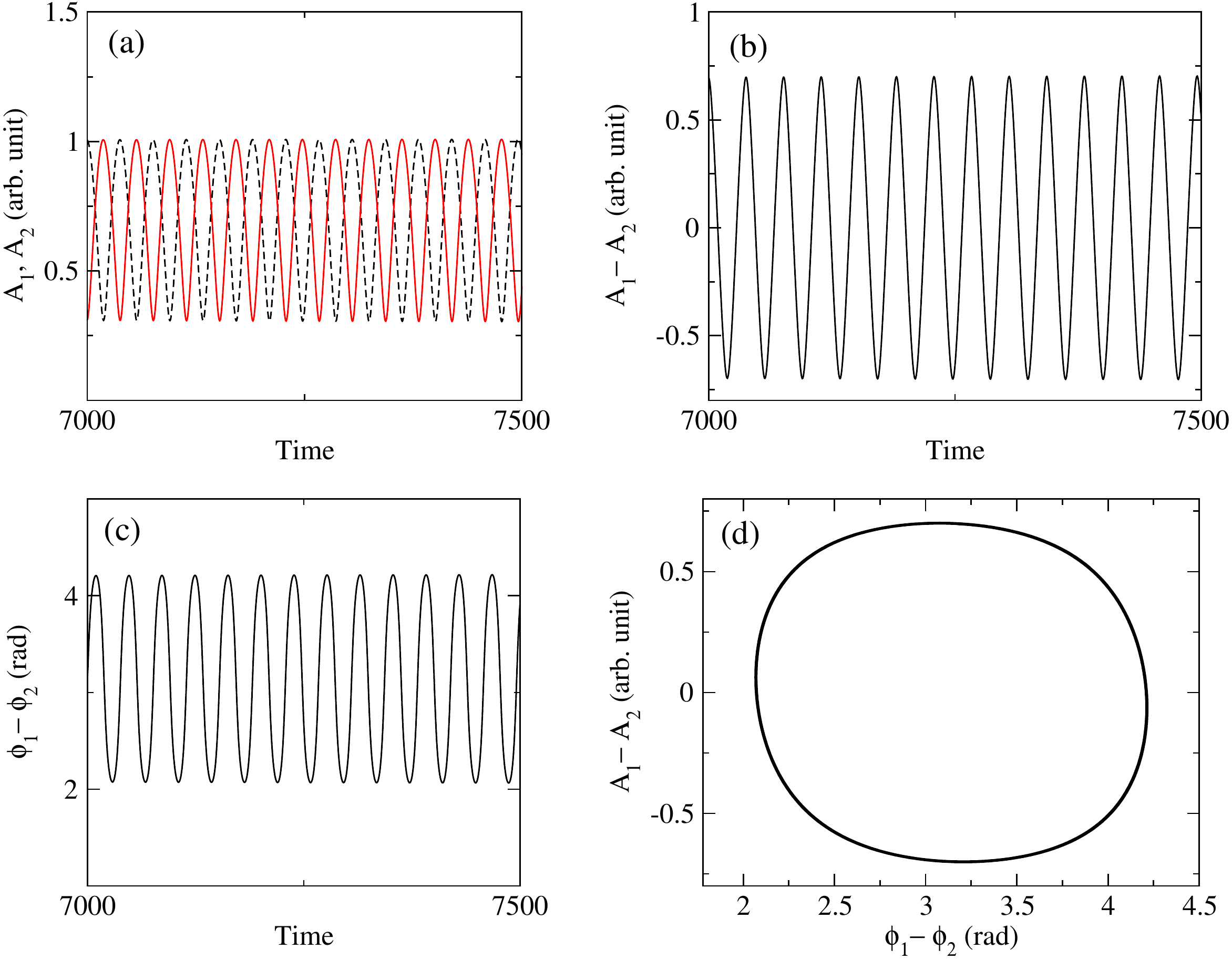}}
\end{center}
\caption{\scriptsize Plots of (a) laser field amplitudes $A_{1}$ and $A_{2}$ (continuous and dashed lines) versus time (in units of cavity photon lifetime), (b) magnified $A_1 - A_2$ versus the same time, (c) $\phi_1 - \phi_2$ versus the same time, and (d) $A_1 - A_2$ versus $\phi_1 - \phi_2$ in the same time span, showing synchronization in the anti-phase periodic state at $\alpha$ = 3.8, for a time delay $\tau=14$ units and a coupling strength $\eta=0.2$.}
\label{sync-periodic}
\end{figure}

Within the window between two successive amplitude-death regions, say at symbol Y in figure \ref{corr1} at $\alpha=3.80$, we have anti-phase periodic oscillations shown in figure \ref{sync-periodic}. Here, the amplitudes and the phases of the laser fields are synchronized, as the oscillations in $A_1 - A_2$ and $\phi_1 - \phi_2$ remain within fixed bounds, around constant average values (figures \ref{sync-periodic}(b) and (c)). The bounded synchronization is evidenced in figure \ref{sync-periodic}(d).

\begin{figure}[htbp]
\begin{center}
\scalebox{0.55}{\includegraphics{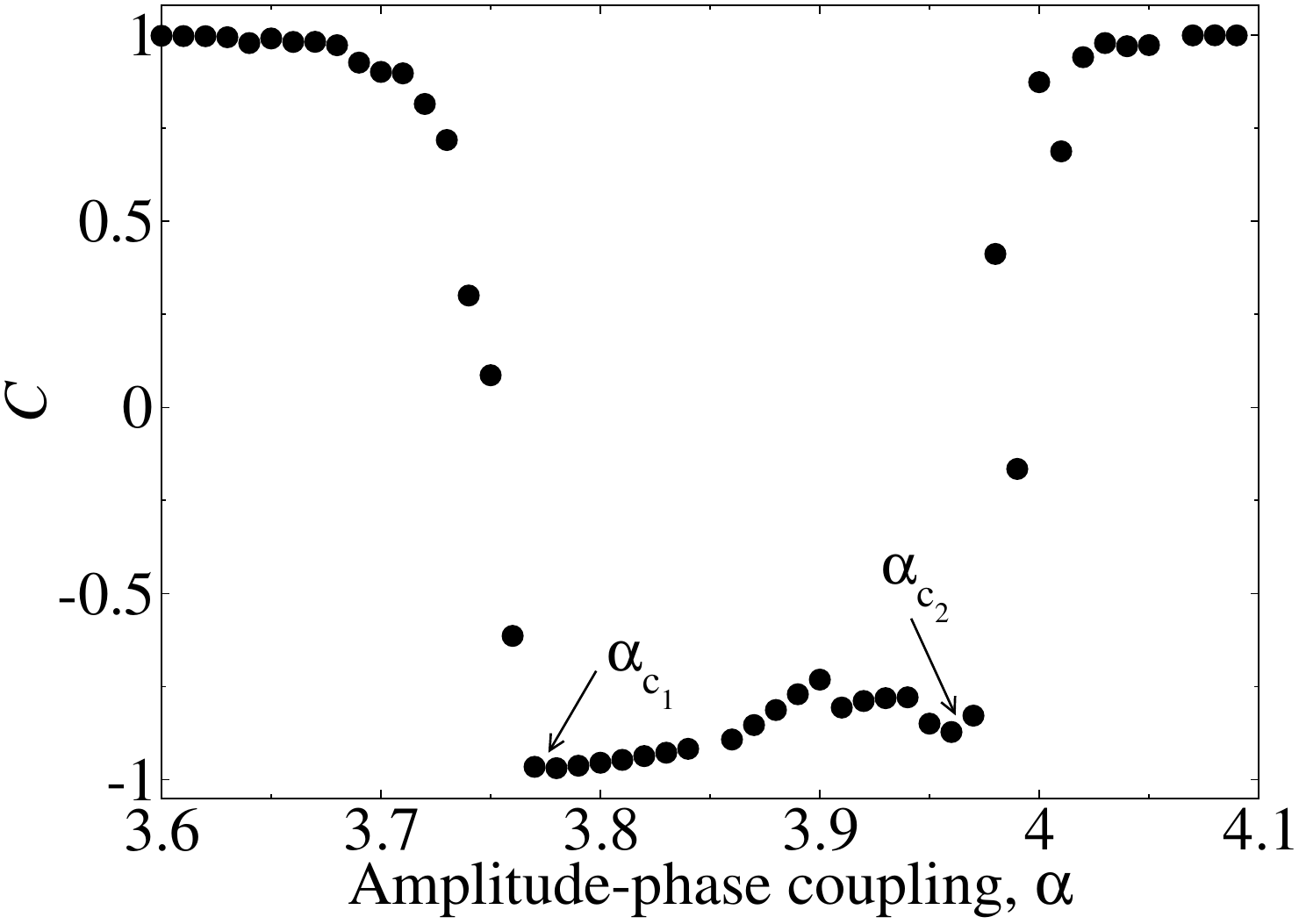}}
\end{center}
\caption{\scriptsize $C$ versus expanded $\alpha$ for one of the windows between successive amplitude-death regions shown in figure \ref{corr1}. $\alpha_{c_{1}}$ and $\alpha_{c_{2}}$ are the critical values of $\alpha$ at which the transitions occur from in-phase amplitude-death to anti-phase periodic oscillations, and vice versa.}
\label{corr2}
\end{figure}

\begin{figure}[htbp]
\begin{center}
\scalebox{0.55}{\includegraphics{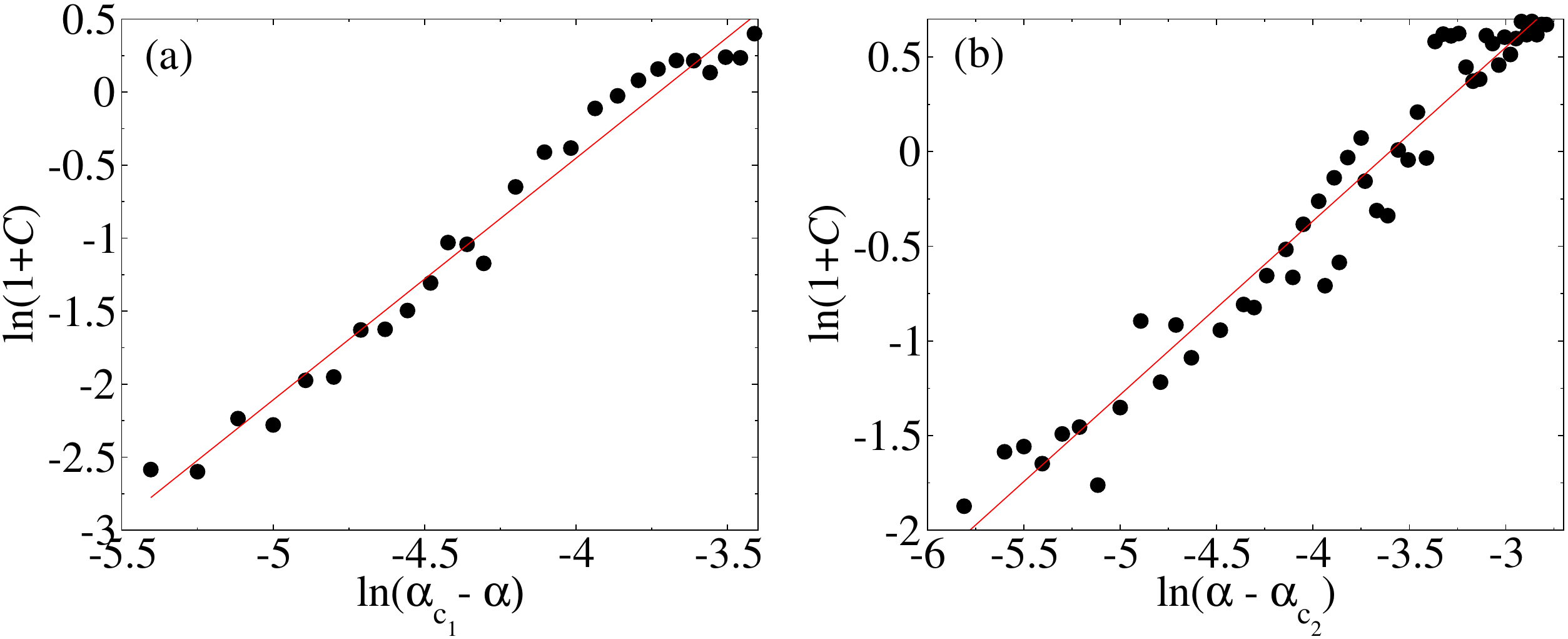}}
\end{center}
\caption{\scriptsize Plot of ln($1+C$) versus ln$|\alpha - \alpha_{c_{1,2}}|$ demonstrating the scaling behaviour of $C$ with respect to (a) $\alpha_{c_{1}}$ and (b) $\alpha_{c_{2}}$ of figure \ref{corr2}. Dots are the numerical data and the solid lines are straight line fittings of the power law (\ref{eq15}). The critical values of the amplitude-phase coupling are found to be $\alpha_{c_{1}} \approx 3.7725$ and $\alpha_{c_{2}} \approx 3.9630$.}
\label{scaling}
\end{figure}

In order to understand the transition from amplitude-death to periodic oscillations and vice versa, we explore one typical window between two successive flat in-phase amplitude-death regions in figure \ref{corr1}. The variation of the cross-correlation coefficient $C$ with the amplitude-phase coupling $\alpha$ in such a window is shown in figure \ref{corr2}. Marked $\alpha_{c_{1}}$ and $\alpha_{c_{2}}$ are the two critical values of $\alpha$ at which the cross-correlation coefficient $C$ indicates transitions from in-phase amplitude-death to anti-phase periodic oscillations and vice versa. To characterize these transitions, we look for a scaling behaviour of $C$ with $\alpha$, as $C$ varies sharply from $1$ to $-1$ and $-1$ to $1$ in figure \ref{corr2}. A power-law behaviour is found as
\begin{equation}
(1+C) \propto |\alpha_{c}-\alpha|^{\mu} , \label{eq15}
\end{equation}
where $\mu$ is the scaling exponent. The fitting of this scaling relation, shown in figures \ref{scaling}(a) and \ref{scaling}(b), gives the exponents $\mu=1.6$ and $0.95$, with $\alpha_{c} \approx 3.7725$ and $3.9630$, respectively, for the two transitions. As mentioned before, $\alpha$ is different for different laser structures, and can be tuned by the carrier density or the light wavelength or the operating temperature to observe this critical behaviour. This general analysis of the complex dynamics can be extended to higher values of $\alpha$, say, 4 to 11, suitable for quantum well devices \cite{markus03}.

\section{Conclusions}
We have explored the role of optical phase dynamics in amplitude and power synchronization in a mutually delay-coupled diode laser system. It is found that the coupling $\alpha$ between the amplitude and the phase of laser fields indeed plays an important role in synchronization of such systems. The optical phase synchronization precedes amplitude-death synchronization, and it drives the field amplitudes to the synchronized death state. The approach to phase synchronization of the oscillating powers of the two coupled lasers can be easily observed experimentally at the appropriate time scale. The phase dynamics of a laser field, on the other hand, cannot be measured directly in an experiment -- the phase difference between the fields can be revealed in the interference of the two fields. Our numerical prediction is expected to prompt experimental tests on such a system.

We have also found that with the increase in amplitude-phase coupling $\alpha$, the system hops over a sequence of in-phase amplitude-death regions. Within the windows between successive in-phase amplitude-death regions, the cross-correlation between the field amplitudes exhibits a power-law behaviour with respect to $\alpha$, unveiling a remarkable universal feature in the dynamics of different kinds of diode lasers.

\ack
VP and AP thank Council of Scientific and Industrial Research, India and Department of Science and Technology, Government of India, respectively, for financial supports. We acknowledge computations performed at the UGC-DRS computing facility at the School of Physical Sciences, Jawaharlal Nehru University.

\end{document}